\documentclass{article}
\usepackage{spconf,amsmath,graphicx}
\usepackage{multirow}
\usepackage{svg}
\usepackage{lipsum}
\usepackage{mathtools}
\usepackage{subcaption}

\graphicspath{{figures/}}
\svgpath{{figures/}}

\DeclareMathOperator*{\argmin}{arg\,min}

\title{Neural Ambisonics encoding for compact irregular microphone arrays}
\twoauthors
  {Mikko Heikkinen}
	{Nokia Technologies\\
	Tampere, Finland}
  {Archontis Politis, Tuomas Virtanen}
	{Tampere University\\
	Tampere, Finland}
\begin{document}
%
\maketitle
\begin{abstract}
Ambisonics encoding of microphone array signals can enable various spatial audio applications, such as virtual reality or telepresence, but it is typically designed for uniformly-spaced spherical microphone arrays. This paper proposes a method for Ambisonics encoding that uses a deep neural network (DNN) to estimate a signal transform from microphone inputs to Ambisonics signals. The approach uses a DNN consisting of a U-Net structure with a learnable preprocessing as well as a loss function consisting of mean average error, spatial correlation, and energy preservation components. The method is validated on two microphone arrays with regular and irregular shapes having four microphones, on simulated reverberant scenes with multiple sources. The results of the validation show that the proposed method can meet or exceed the performance of a conventional signal-independent Ambisonics encoder on a number of error metrics.

\end{abstract}

\section{Introduction}
\label{sec:introduction}

Ambisonics is a device-independent representation for spatial audio capture and reproduction that has been adopted widely by industry for immersive media and VR/AR/XR applications \cite{zotter2019ambisonics,olivieri2019scene}. Transforming microphone array signals to a device-independent representation keeps the device-specific parts of the processing to a minimum. This means that it is easier to build a scalable solution for spatial audio processing and delivery with, e.g., source separation or beamforming taking place on the Ambisonic signals. Also, Ambisonics can be used as a spatial audio transport and storage format \cite{nachbar2011ambix,herre2015mpeg}. 

Many customer devices, such as mobile phones and head-mounted displays, already have microphone arrays to enable use cases unrelated to spatial audio capture, e.g., acoustic echo cancellation in hands-free communication use. However, such devices are not designed primarily for spatial audio, and their shapes do not allow a favourable placement of microphones uniformly over a sphere. So it is valuable to have a solution to deliver Ambisonics signals from devices having any microphone array shape, to use them more flexibly for various spatial audio applications.

Ambisonic encoding is typically treated as an optimal matching problem, in a least-squares sense, between the directional responses of the array and the directional responses of the Ambisonic signals, the spherical harmonics (SHs), which simplifies considerably in the case of SMAs \cite{zotter2019ambisonics, jin2013design, zotkin2017incident, politis2017comparing}. A similar least squares solution can be adopted for irregular arrays, resulting in an encoding matrix of filters \cite{laborie2003new, zotkin2017incident, jin2013design, politis2017comparing, bastine2022ambisonics}. With the rise of wearable arrays on XR devices expected to provide immersive audio capturing, the problem of Ambisonic encoding on irregular geometries has attracted increased attention lately \cite{mccormack2022parametric, bastine2022ambisonics, ahrens2022spherical}. In \cite{ahrens2022spherical} the encoding problem is constrained to a 2-dimensional sound field, resulting in perceptually improved encoding at the expense of reduced elevation rendering. The method in \cite{mccormack2022parametric} proposes a signal-dependent encoder assuming a primary-ambience directional model, in order to overcome limitations of signal-independent encoding. However, its performance relies on the accurate estimation of model parameters.

Very few works have focused on learning-based Ambisonics processing. Gao et al. \cite{gao2022sparse} aim to improve encoding performance for higher-order SMAs, such as the 32-capsule Eigenmike, using a convolutional neural network (CNN). Their focus is especially above the spatial aliasing frequency limit of the SMA where conventional encoding deteriorates. More recently, \cite{zhu2022binaural} proposed a DNN-based Ambisonics to binaural decoder, based on a U-Net model and combining time- and frequency domain losses in the training. A related work by Hsu et al. \cite{hsu2023model} uses a model matching principle to design a learnable U-Net based encoder-decoder that renders microphone array recordings to binaural signals. Even though different from our target, the transformation is similar enough that the results may apply also to the proposed domain.

This paper proposes a deep learning-based Ambisonics encoding method that is not constrained to special geometries. Similar to \cite{gao2022sparse,zhu2022binaural} it uses a U-net structure to estimate a complex encoding matrix, with the addition of a frequency-specific preprocessing layer that improves model performance compared to using only a U-Net. We also propose a novel loss function that introduces energy preservation and coherence components to condition the model.
The model is evaluated in conditions with a variable number of sound sources and varying degrees of reverberation to make the problem more realistic and challenging.
Evaluation is carried out against a traditional Ambisonics encoder consisting of a matrix of time-invariant filters. The evaluation proves the proposed method to be a promising approach with performance meeting or exceeding that of the baseline method on the selected metrics.

\section{Proposed Methods}
\label{sec:methods}

\subsection{Problem Setup}
\label{ssec:problem_setup}

The sound field in Ambisonics is modeled as the spherical harmonic (SH) transform of the continuous distribution of plane wave amplitudes expressing the sound scene. Following the definitions in \cite{politis2017comparing}, we define the Ambisonics signal set, up to SH (or Ambisonic) order $N$, as
\begin{equation}
\begin{aligned}
\mathbf{b}(t, f)&=\iint a(t,f,\theta,\phi)\mathbf{y}_N(\theta,\phi)d{\theta}d{\phi}.
\end{aligned}
\label{eq:ambisonics_model}
\end{equation}
where $\mathbf{y}_N = [Y_{00},...,Y_{nm},...,Y_{NN}]^\mathrm{T}$ is a vector of real-valued SHs for azimuth $\theta$ and elevation angle $\phi$, indexed by SH order $n$ and degree $m$ up to a maximum encoding order $N$. $\mathbf{b}=[b_{00},...,b_{nm},...,b_{NN}]^\mathrm{T}$ are the $(N+1)^2$ ambisonic signals indexed in the same manner as the SHs and  $a(t, f, \theta,\phi)$ is a plane wave amplitude distribution describing the incident sound field. Assuming time-frequency transformed signals, $t$ is the time and $f$ the frequency index.

Similarly, we define the signals captured by a microphone array of $q=1,...,Q$ microphones, in the same sound field, in terms of the scene plane wave distribution and the array transfer functions (ATFs) $\mathbf{h} = [h_1,...,h_Q]^\mathrm{T}$ as

\begin{equation}
\mathbf{x}(t, f)=\iint a(t,f,\theta,\phi) \mathbf{h}(f,\theta,\phi)d{\theta}d{\phi}.
\label{eq:mic_array_model}
\end{equation}

Our aim is to estimate a complex multichannel encoding matrix $\mathbf{M}(t,f)$ for every time-frequency point that transforms the microphone signals $\mathbf{x}$ to an approximation of the ideal ambisonic signals $\bar{\mathbf{b}} \approx \mathbf{b}$ that are as similar to them according to some optimality criteria, with
\begin{equation}
\bar{\mathbf{b}}(t, f) = \mathbf{M}(t,f) \mathbf{x}(t, f)
\label{eq:estimation_model}
\end{equation}

\subsection{DNN Method}
\label{ssec:dnn_method}

\begin{figure}[]
  \centering
  \includegraphics[width=0.45\textwidth]{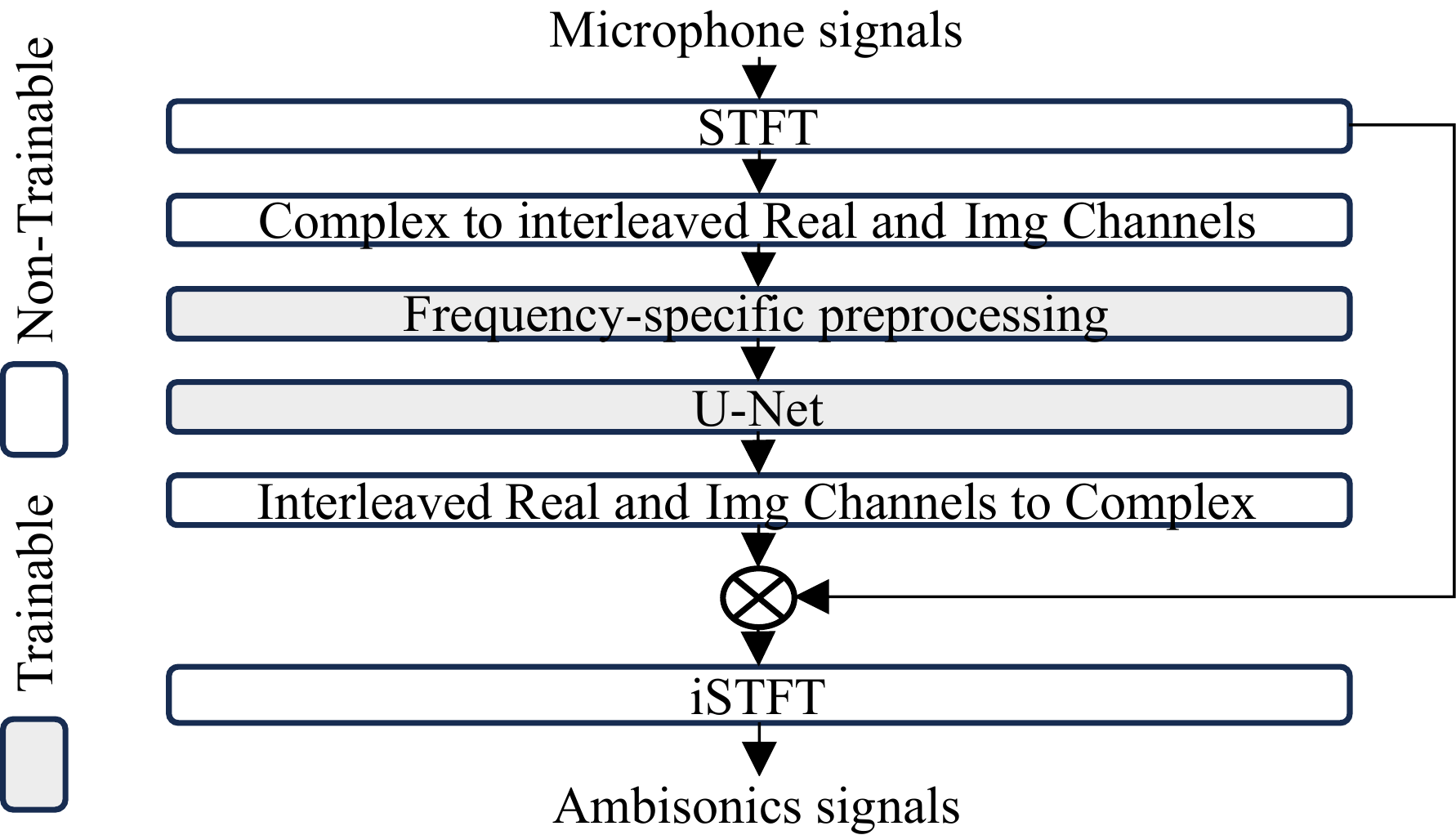}
  \caption{Block diagram of the proposed method.}
  \label{fig:processing_chain}
\end{figure}


A block diagram of the proposed method is presented in Figure \ref{fig:processing_chain}. A short-time Fourier transform (STFT) is applied to all channels of the microphone array. The real and imaginary parts of the STFT data are separated to real-valued feature channels. Next, the data goes through the learnable stages of preprocessing and U-Net. Feature channel pairs are then combined to form a complex-valued encoding matrix which, when multiplied with the input signals in the STFT domain, produces the final Ambisonics signals. 

Sound from sources will reach microphones in the array at different times according to their angle of incidence to the microphone, producing frequency dependent phase differences.
Frequency dependent behaviour is also required for arrays with complex scattering, and frequency dependent directivities. U-Net uses 2D convolution layers, sharing coefficients for all time steps and frequencies, which does not allow modeling of frequency-dependent behaviour. Therefore we have added a separate preprocessing stage to adapt the input prior to the U-Net processing. 

The frequency-specific preprocessing is implemented as a 2D convolution where each frequency bin has unique filter coefficients that are shared for all time steps. The convolutions are applied with a sliding window across time steps so that each convolution sees a 2D local window to the neighboring time and frequency bins as input on each step. The preprocessing is implemented using a 2D locally-connected convolution layer \cite{taigman2014deepface} of size one times the number of frequencies. We experimented with using only the U-Net and replacing the 2D convolutions in the U-Net with layers similar to the proposed frequency-specific preprocessing and found the approach of having a separate preprocessing and an unmodified U-Net to be a better solution.

The U-Net follows the original design described in \cite{ronneberger2015u}. U-Net encoder blocks consist of two consecutive 2D convolutions with nonlinear activation functions followed by batch normalization, dropout, and 2D max pooling layers. Decoder blocks consist of 2D transposed convolution, which is concatenated with a skip connection, followed by a dropout layer and two consecutive 2D convolutions with Rectified linear unit (ReLU) activation functions. The bottom bottleneck layer has two consecutive 2D convolutions with ReLU activations. The skip connections are taken from the encoder layers after the dropout layer before max pooling is applied.

A composite loss function is used in model training. The function consists of a mean absolute error (MAE) component, a coherence component, and an energy preservation component. The components are weighted and applied to specific frequency ranges. The MAE component is applied to the simulated reference $y$ and predicted $\hat{y}$ STFTs as
\begin{equation}
MAE(f)=\frac{1}{TC}\sum_{t=1}^T\sum_{c=1}^C|y_{t,f,c}-\hat{y}_{t,f,c}|.
\label{eq:mae}
\end{equation}
$T$ and $C$ are the number of time steps and microphone channels in the input. Model training with this component alone produces results that are comparable to the baseline model below the spatial aliasing frequency.

The energy preservation component controls the amplitude of high frequency content. The Ambisonics transform preserves the total sound field energy, and that behaviour is desirable even at high frequencies where the encoder has difficulty maintaining the ideal directivity of individual Ambisonic channels. This component is defined as
\begin{equation}
E(f)=\frac{1}{T}\sum_{t=1}^{T}\left (\big|{\sum_{c=1}^{C}|y_{t,f,c}|^2-\sum_{c=1}^{C}|\hat{y}_{t,f,c}|^2}\big|\right ).
\label{eq:loss}
\end{equation}
The coherence component is defined as
\begin{equation}
C(f) = \frac{1}{C}\sum_{c=1}^C\frac{|\sum_{t=1}^{T} y^*_{t,f,c} \hat{y}_{t,f,c} |^2 }{ \sum_{t'=1}^T |y_{t',f,c}|^2 \sum_{t''=1}^T |\hat{y}_{t'',f,c}|^2 }
\label{eq:coherence}
\end{equation}
where $y^{\ast}$ is the complex conjugate of the input signal. The coherence component aims to capture how well the directivity patterns of the Ambisonics channels are reproduced.

The final loss function is the weighted sum of the above losses, where the weights are frequency dependent:
\begin{equation}
\begin{aligned}
L= \frac{1}{F}\sum_{f=1}^F\left( \alpha(f){\cdot}MAE(f)+\beta(f){\cdot}E(f)+\gamma(f){\cdot}C(f)\right ).
\label{eq:loss}
\end{aligned}
\end{equation}
Above, $F$ is number of frequency bins in the input. The $\alpha(f)$, $\beta(f)$, and $\gamma(f)$ are frequency-dependent weights that control the range on which each loss component is applied.

\section{Evaluation}
\label{sec:evaluation}

\subsection{Data}
\label{ssec:data}


\begin{figure*}[t]
  \centering
  \includegraphics[width=\textwidth]{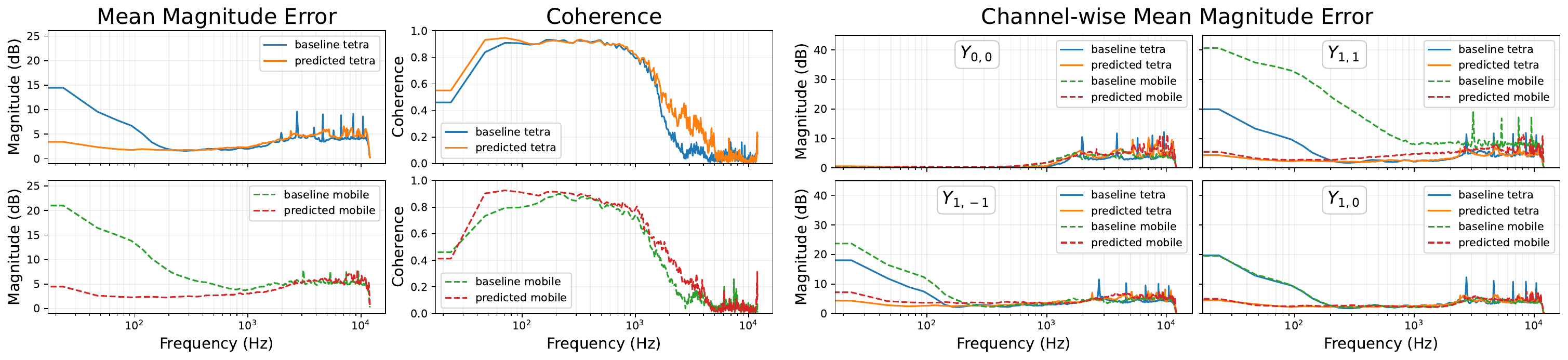}
  \caption{On the left: evaluation metrics averaged over all channels. On the right: mean magnitude errors for individual channels.}
  \label{fig:combined_results_plot}
\end{figure*}

Audio scenes of a microphone array placed in a room capturing audio from one to three simultaneous sound sources were simulated. The room sizes were randomized between [3, 8], [3,12], [3, 20] meters height, width, and depth. Source locations in the room were randomized. Minimum distance of microphone array to any wall was 1 m, and minimum distance of a sound source to the microphone array was 2 m.

Two microphone array geometries were used in evaluation. The first was a slim rectangular shape reminiscent of a mobile phone. Two microphones were situated in each end of the shape creating an irregular array with a large difference between the smallest and largest microphone distances, with coordinates in meters [[0.08, 0, 0.03], [0.08, 0, -0.03], [-0.08, 0.005, 0.005], [-0.08, -0.005, -0.005]]. The second was a regular tetrahedron geometry of radius 9 cm. The size of the tetrahedron array was chosen to be similar to the first array. The microphone array was simulated as an acoustically transparent one of ideal omnidirectional microphones. An omnidirectional assumption simplifies simulations. A real-world array such as a mobile phone with multiple microphones will have direction-dependent characteristics caused by scattering of the device body and the microphone mountings. 

Multichannel room impulse responses (RIRs) were simulated with the image-source method using Pyroomacoustics package \cite{scheibler2018pyroomacoustics} and the source signals were convolved with them to produce the array recordings. Ambisonic reference signals were also simulated at the center point of the microphone array by creating Ambisonic RIRs from the image source model. The impulse response was transformed to the Ambisonics domain with the spherical harmonic transform (SHT) considering each image source as a far-field point source. This method avoids spatial aliasing. The input source signal was convolved with the Ambisonics RIR to produce the reference signal for that source.

Simulations of 10,000 2 second sound scenes were created, and divided into training, validation, and evaluation datasets with a 80/10/10 split. The ESC-50 dataset \cite{piczak2015esc} provided source signal data. The source samples were divided randomly to the above splits. The sound scenes were randomly generated for the datasets, so it is unlikely that there were scenes with the exact same acoustic setups.

\subsection{Model parameters}

Sampling rate of 24 kHz was used. STFT used FFT length of 1024 with 512 hop size and Hann window.
The model input size was 94x513 [$T$, $F$]. The preprocessing had $F$ 3x3 Conv2D operations with 8 input and output channels. U-Net input size was zero padded to 96x560. Each encoder layer halved time and frequency dimensions. The number of output channels were {32, 64, 128 ,256}, 512, and {256, 128, 64, 32}, for encoder, bottleneck, and decoder layers, respectively. Convolution kernel sizes were 3x3 in all layers. Dropout probabilities were 0.25 for encoder and 0.5 for decoder layers. ReLU activations were used in all layers. 

The loss parameters were $\alpha$: 1.0 from 0 to 1 kHz, linearly ramped from 1.0 to 0 between 1 and 2 kHz, and 0 from 2 to 12 kHZ. $\beta$: 0 from 0 to 1 1Khz, linearly ramped from 0 to 0.01 between 1 and 2 kHz, and 0.01 from 2 to 12 kHz. $\gamma$: 5.0 from 0 to 4kHz, linearly ramped from 5.0 to 0 between 4 and 5kHz, and 0 from 5 to 12 kHz. The model hyperparameters and loss parameters were selected experimentally to minimize validation loss. Training was carried out with Adam optimizer, with 5e-5 learning rate, and batch size of 32.

%

\subsection{Baseline}
\label{ssec:baseline}

For the baseline method we use a conventional time-invariant encoding matrix $\mathbf{M}(f)$ which comes as the solution to the following least squares objective
\begin{equation}
    \mathbf{M}_\mathrm{b}(f) = \argmin_\mathbf{M} \iint ||\mathbf{M}(f)\mathbf{h}(f,\theta,\phi) - \mathbf{y}_N(\theta,\phi)||^2\mathrm{d}\theta\mathrm{d}\phi.
\end{equation}
The solution can be derived by discretizing the integral at a number of points on the sphere, or by employing the SH transform of the ATFs \cite{jin2013design,politis2017comparing} and can be applied to modeled or measured ATFs. Typically, Tikhonov regularization is introduced in the solution, to avoid excessive filter amplification at low frequencies for Ambisonic channels of $n\geq1$. Above the spatial aliasing frequency limit the Ambisonic channels do not maintain anymore the intended SH directivities. On that range a diffuse-field equalization is performed, as proposed in \cite{gerzon1975design} to combat a perceived high frequency colouration. In this work, we apply regularization that limits filter amplification to 15dB and diffuse-field equalization as in \cite{mccormack2022parametric}.

\subsection{Metrics}
\label{ssec:metrics}

Spectral errors were evaluated with a mean magnitude spectrum error across individual channels:
\begin{equation}
S(f)=\frac{1}{TC}\sum_{c=1}^{C}\sum_{t=1}^{T}\left|20\cdot log_{10}\left( \frac{|y_{t,f,c}|}{|\hat{y}_{t,f,c}|}\right)\right|
\label{eq:magnitude_spectrum_err}
\end{equation}
To evaluate also spatial quality of encoding, the magnitude squared coherence of Eq. (\ref{eq:coherence}) is also used as a metric. 

\subsection{Results}
\label{ssec:results}

Figure \ref{fig:combined_results_plot} shows the evaluation metrics for both microphone arrays plotted against frequency. Overall, the proposed method can meet or exceed the performance of the baseline model in both metrics.
The conventional baseline encoder has an optimal frequency range of about 200 Hz -- 700 Hz for the tetrahedral array, below which the filters do not counteract effectively anymore the energy loss of the first-order components, causing an increase of spectral errors. Above that range and the spatial aliasing limit of the array, the baseline also fails to maintain the proper omnidirectional and dipole directivities of the Ambisonic channels, causing the increased loss of spatial coherence. In addition, since the array is an open design, without any scatterers involved such as a mounting baffle, there are large error spikes in all channels due to severe frequency dips not recovered effectively by the encoding filters; a well known effect of open SMAs \cite{balmages2007open}. The situation is even worse for the irregular array, since no usable frequency range across all channels can be observed where encoding achieves both low spectral error and high spatial coherence simultaneously. Looking at spectral errors of the individual channels, it is evident that the channel aligned with the larger spacing of microphones ($Y_{1,1}$) suffers the worst performance, while the rest of the channels seem to have a performance comparable to the tetra array.

The proposed signal-dependent encoder manages to achieve a reasonable performance for all Ambisonic components at low frequencies, in terms of spectral differences, without significant differences between the regular and irregular array. More importantly, the method seems to be extending the usable frequency range above the aliasing limit to some degree, especially for the irregular array, as indicated by the mean spatial coherence plot across channels. This is an interesting result and indicates a direction for further research, since aliasing is a hard limitation for conventional encoding filters. Furthermore, error spikes at high frequencies for both arrays are absent with the proposed method, showing that they can be counteracted with time-variant signal dependent encoding.





\section{Conclusions}
\label{sec:conclusion}

In this work, we proposed a machine-learning solution for Ambisonics capture on a small and possibly irregular shaped microphone array. A U-Net architecture was adapted to the problem by adding learnable preprocessing and novel loss function combining MAE, coherence, and energy preservation components. Evaluation against a linear Ambisonics encoder shows that the method can achieve smaller magnitude spectrum error in frequencies below spatial aliasing, especially on an irregularly shaped array, and slightly better coherence above spatial aliasing frequency.

Evaluation in a setting that introduces microphone directivities and scattering caused by device enclosure  is an interesting topic for further research. Training and test signals with similar characteristics were used in this a proof of concept. Future work also includes investigating more diverse material and the effect of training-test mismatches.

 
 
 
 

\vfill\pagebreak

\bibliographystyle{IEEEbib}
\bibliography{refs}

\end{document}